\documentclass[aip,pop,reprint]{revtex4-1}
 \usepackage{dcolumn}
 \usepackage{bm}
 \usepackage{amsmath}

\newcommand{\DL}{\textrm{DL}}
\newcommand{\erfc}{\textrm{erfc}}
\newcommand{\p}{\textrm{P}}

\newcommand{\vc}{\mathbf}
\usepackage{graphicx}

\begin{document}


\title{Particle-in-cell simulations of a current-free double layer}


\author{S.\ D.\ Baalrud\footnote{Present address: Center for Integrated Computation and Analysis of Reconnection and Turbulence, University of New Hampshire, Durham, New Hampshire 03824}}
\author{T.\ Lafleur}
\author{R.\ W.\ Boswell}
\author{C.\ Charles}
\affiliation{Space Plasma, Power and Propulsion Group, Research School of Physics and Engineering, The Australian National University, Canberra ACT 0200, Australia }

\date{\today}

\begin{abstract}

Current-free double layers of the type reported in plasmas in the presence of an expanding magnetic field [C.\ Charles and R.\ W.\ Boswell, Appl.\ Phys.\ Lett.\ {\bf 82}, 1356 (2003)] are modeled theoretically and with particle-in-cell/Monte Carlo simulations. Emphasis is placed on determining what mechanisms affect the electron velocity distribution function (EVDF) and how the EVDF influences the double layer. A theoretical model is developed based on depletion of electrons in certain velocity intervals due to wall losses and repletion of these intervals due to ionization and elastic electron scattering. This model is used to predict the range of neutral pressures over which a double layer can form and the electrostatic potential drop of the double layer. These predictions are shown to compare well with simulation results. 



\end{abstract}




\maketitle


\section{Introduction}

Double layers are adjacent regions of net positive and negative charge that form distant from the physical boundaries of a plasma.  They typically provide an electrostatic boundary that separates plasmas with different properties. There are several varieties of double layers,\cite{hers:85,char:07} some of which have been studied since the earliest days of plasma physics research.\cite{lang:29} One categorization is that double layers can be either current-carrying or current-free.  The current-free variety was predicted theoretically in the early 1980s,\cite{perk:81} and these were later observed experimentally.\cite{hata:83,hair:90} Recently, a renewed interest in current-free double layers\cite{char:03,cohe:03,sun:04,sun:05,kees:05,suth:05,plih:07,chak:09,bilo:09,chak:10,scim:10,lafl:10,fred:10} has arisen in part because of their application to electrostatic thrusters for spacecraft propulsion\cite{char:06,char:09b,west:09,ling:10} and auroral physics.\cite{char:09c}

\begin{figure}
\includegraphics{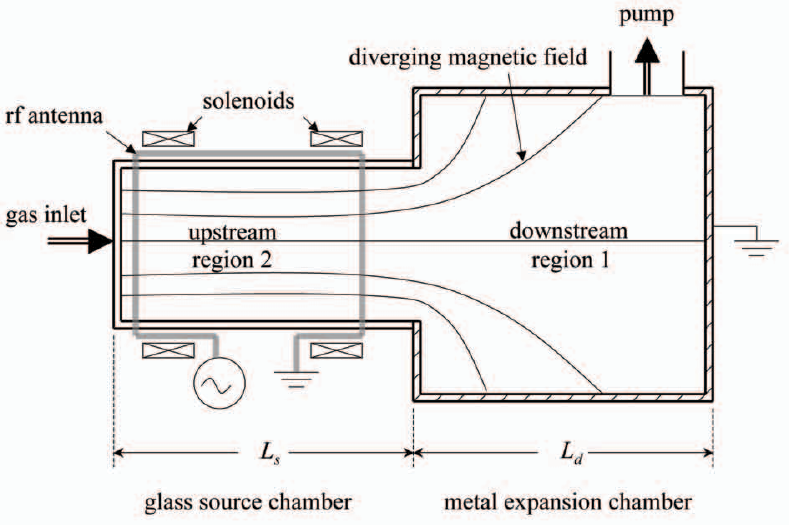}
\caption{Schematic diagram of a common experimental apparatus for studying current-free double layers in an expanding magnetic field. }
\label{fg:diagram}
\end{figure}

These recent current-free double layer experiments\cite{char:03,cohe:03,sun:04,sun:05,kees:05,suth:05,plih:07,chak:09,bilo:09,chak:10,scim:10,lafl:10,fred:10} consist of an insulated source chamber connected to a larger volume expansion chamber that is metallic and grounded; see Fig.~\ref{fg:diagram}. An approximately constant axial magnetic field is applied to the source chamber, which diverges near the boundary between the source and expansion chambers.  Plasma is generated in the source chamber by applying rf waves with an antenna.  Current-free double layers have been measured in the region of divergent magnetic field in this configuration.\cite{char:03,cohe:03} It has also been confirmed that these double layers generate an ion beam in the expansion chamber that has a flow speed typically a few times faster than the ion sound speed.\cite{sun:05,kees:05}   

Analytic models of current-free double layers in expanding plasmas have been proposed by Chen,\cite{chen:06} Lieberman \textit{et al},\cite{lieb:06} Goswami \textit{et al},\cite{gosw:08} and Ahedo and S\'{a}nchez.\cite{ahed:09} These are fundamentally different in that each makes a different assumption for the electron velocity distribution function (EVDF). Chen considers just the upstream region and assumes that electrons are Maxwellian.\cite{chen:06} Lieberman \textit{et al} consider two populations of electrons upstream: a thermal (Maxwellian) population and an additional half-Maxwellian beam population.\cite{lieb:06} The upstream electrons in Goswami \textit{et al} are counter-streaming Maxwellian beams.\cite{gosw:08} Ahedo and S\'{a}nchez assume a two-temperature Maxwellian distribution characterized by hot and cold populations.\cite{ahed:09} Double layer formation is sensitive to the EVDF, so each of these theories predicts different double layer parameters such as the potential drop and resultant ion beam properties.

An accurate model of the EVDF, and experimental verification of it, is needed to provide a foundation for a comprehensive analytic model of the experiments. In particular, verification of the electron beams assumed to be present in the source chamber in Refs.~\onlinecite{lieb:06} or \onlinecite{gosw:08} is lacking. Unfortunately, diagnosing the EVDF is difficult to do experimentally. Essentially the only diagnostic available is a Langmuir probe, but this is typically limited to measuring the electron energy distribution function (EEDF) rather than the EVDF. Another limitation of Langmuir probes is that current-voltage characteristics get noisy for energies greater than a couple of electron temperatures.  Previous measurements have given ambiguous results concerning electron beams in the source region. Early work with Langmuir probes provided some ``preliminary'' evidence of an electron beam very close to the sheath of the source chamber.\cite{char:04}  Other indirect measurements associated with an ionization instability also appeared to suggest that electron beams were present.\cite{aane:06} 
However, more recent Langmuir probe measurements have found no evidence of beams.\cite{taka:07,taka:09,taka:10a} Instead, these found a Maxwellian EEDF that was depleted in density beyond the double layer potential energy. 

In this work, we develop a model for the EVDF in an expanding plasma and compare the results with PIC simulations. We concentrate on a simplified geometry that has only one spatial dimension, but is 3D in velocity phase-space. The boundary conditions on the geometric dimension are an insulating wall at one end (source chamber) and a conducting wall at the other (expansion chamber). The analytic model accounts for depletion of velocity phase-space intervals due to loss of electrons to the boundaries, as well as partial repletion of these intervals due to ionization sources and scattering.  The model predicts that a current-free double layer can only exist over a finite range of neutral pressures. It can also be used to predict the double layer and sheath potential drops based on the electron temperature. 

The PIC code, named {\sc phoenix}, uses the same 1D in space, 3D in velocity phase-space geometry that the analytic model is based on.  Collision processes are modeled using a Monte Carlo algorithm and energy is input with a method that simulates inductive electron heating in a velocity-space direction perpendicular to the geometric domain.  Plasma expansion is modeled by invoking a loss profile in the downstream region. Aside from details of the loss profile used, {\sc phoenix} has been designed to be identical to the code developed by Meige \textit{et al}.\cite{meig:05,meig:05b,meig:06} We find that the EVDFs calculated using the PIC code differ substantially from those assumed in previous literature.\cite{chen:06,lieb:06,gosw:08,ahed:09} Electron beams are not observed, which is consistent with the most recent Langmuir probe measurements.\cite{taka:07,taka:09,taka:10a} The EVDFs in the simulations are shown to agree with the analytic model based on depletion due to wall losses and partial repletion due to scattering.

This paper is organized as follows. Section~\ref{sec:basic} develops a model for the EVDF starting from academic examples which highlight the physics behind the maximum double layer potential drop, as well as the minimum and maximum neutral pressures that can support a double layer. A model of the whole 1D simulation domain is given in Sec.~\ref{sec:full}. After describing the {\sc phoenix} code in Sec.~\ref{sec:code}, the simulation results are provided in Sec.~\ref{sec:sims}. This section also contains a discussion of how the simulated EVDFs relate to previous work and compare with the analytic model of Sec.~\ref{sec:basic}. The results are summarized in Sec.~\ref{sec:sum}. 

\section{Model of double layer formation and the EVDF\label{sec:basic}}

\begin{figure}
\includegraphics{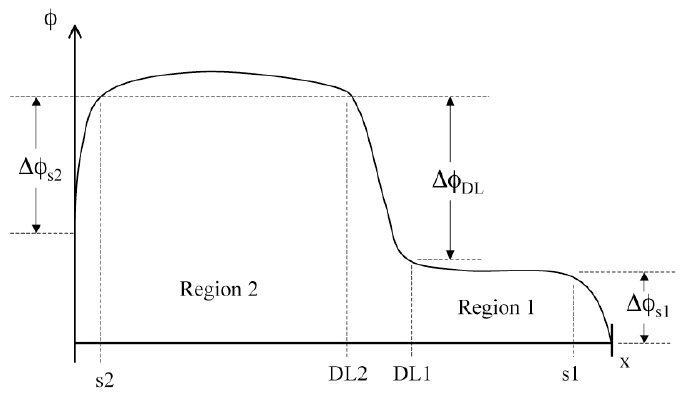}
\caption{Sketch of a typical potential profile for a current-free double layer in the experimental geometry shown in Fig.~\ref{fg:diagram}.}
\label{fg:phi_sketch}
\end{figure}

The electrostatic potential profile along the axis of previous current-free double layer experiments is shown schematically in Fig.\ \ref{fg:phi_sketch}.\cite{char:03,cohe:03} The boundary on the upstream side (region 2) of the experiments is insulating, while the downstream boundary of the larger expansion chamber is conducting and grounded, which we take to be the reference potential. Of course, the experiments, which are cylindrical, have radial profiles in the transverse direction that can affect the details of the axial profile at different radial positions.\cite{corr:08,taka:08,taka:10c} Although consideration of these 3D affects are necessary in order to quantitatively model the experiments, we consider a simplified 1D model here. Our goal is to identify what mechanisms influence the EVDF and, as a result, the double-layer and sheath potentials. The simulations presented in Sec.\ \ref{sec:sims} also use the 1D geometry and thus provide a proving ground for comparison to this model. 

Since the upstream boundary is insulating, it must collect equal fluxes of electrons and ions (assumed to be singly charged here) during steady-state operation. The only other physical boundary in this system is the grounded downstream wall, thus it too will collect equal electron and ion currents (we assume no external electron or ion sources, the only source is ionization which produces electrons and ions in equal numbers). In the absence of any current sources or sinks, a consequence of the current-free boundary conditions is that the double layer must also be current-free. If the EVDF [$f_{e,x}(v_x)$] is known at the positions P = S2, DL2 and S1, denoting the upstream sheath edge, the upstream double-layer edge, and the downstream sheath edge respectively, the current-free condition
\begin{equation}
\int_{-\infty}^{\infty} dv_x\, v_x\, f_{e,x} (x = \textrm{P}) = e^{-1/2} n_{\textrm{P}} c_{s,\textrm{P}} \label{eq:curbal}
\end{equation}
can be used at each of these locations to determine the upstream sheath potential drop ($\Delta \phi_{s2}$), the double layer potential drop ($\Delta \phi_{\DL}$) and the downstream sheath potential drop ($\Delta \phi_{s1}$). The right side of Eq.\ (\ref{eq:curbal}) is the Bohm flux for ions and $c_{s} \equiv \sqrt{T_e/M_i}$ the ion sound speed. The $e^{-1/2}$ term is from the density drop caused by the presheath. 

In the following four sections, we develop a model for $f_{e,x}$ that can be used in Eq.~(\ref{eq:curbal}) to calculate the double layer potential drop as well as determine a neutral pressure range that can support it. Section~\ref{sec:chen} starts with a simplified geometry in which both regions 1 and 2 are semi-infinite domains. This geometry, which has also been studied by Chen,\cite{chen:06} provides a maximum double layer potential drop. The following sections, \ref{sec:pmin} and \ref{sec:pmax}, account for the upstream and downstream walls, which leads to predictions for the minimum neutral pressure and maximum neutral pressure that can support a double layer. Section~\ref{sec:full} puts these geometries together to form a comprehensive model for the EVDF that accounts for both upstream and downstream boundaries. 

\subsection{Semi-infinite domains: $|\Delta \phi_{\DL}|_{\max}$\label{sec:chen}}

We start with perhaps the simplest conceptual current-free double layer configuration.  Here it is assumed that plasma is generated in an upstream source region that is sufficiently large that the plasma has a nominal Maxwellian distribution (i.e., the source chamber is longer than either the electron-electron or electron-neutral collision length). The downstream region is assumed to be infinite and collisionless, so all particles that escape the source region remain downstream. This configuration may be relevant to a thruster operating in space, where the thruster is the source and downstream is space vacuum. 

\begin{figure}
\includegraphics{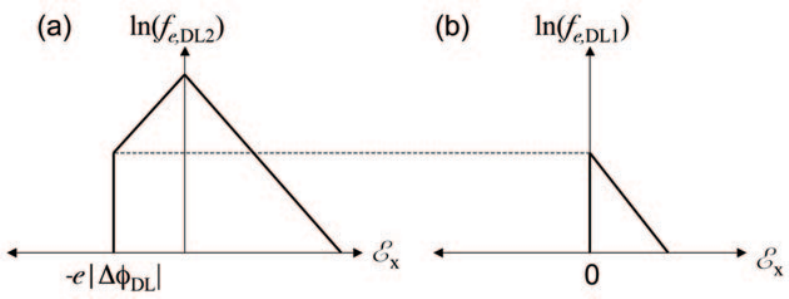}
\caption{Sketch of the natural log of the EVDF for the semi-infinite domains of Sec.~\ref{sec:chen} at positions (a) just upstream of the double layer (P=DL2), and (b) just downstream of the double layer (P=DL1).}
\label{fg:fea_sketch}
\end{figure}

The expected EVDF just upstream and downstream of the double layer is shown in Fig.~\ref{fg:fea_sketch} for this configuration. Here $\mathcal{E}_x \equiv \frac{1}{2} m_e v_x |v_x|$ is an energy variable that accounts for the particle direction. At position DL2 (just upstream of the double layer), the distribution is Maxwellian in the velocity interval $v_x \geq 0$, which consists of thermal electrons migrating from the upstream region. It is also Maxwellian in the interval $-v_{\DL}\equiv - \sqrt{2 e|\Delta \phi_\DL|/m_e} \leq v_x \leq 0$, which consists of thermal electrons from the source that were subsequently reflected from the double layer electric field. The distribution is empty in the interval $v_x \leq -v_{\DL}$ since these electrons had enough directed energy to traverse the double layer and escape downstream. Downstream, these electrons create a half-Maxwellian distribution. The EVDF at position DL2 can thus be written
\begin{eqnarray}
f_{e,\DL2} = \frac{e^{- v_x^2 / v_{Te}^2}}{\sqrt{\pi} v_{Te}}  \label{eq:fa}
\left\lbrace \begin{array}{cc}
0, & v_x < -v_{\DL} \\
n_{a,\DL2} , & -v_{\DL} \leq v_x
\end{array} \right.  .
\end{eqnarray}
Here $n_{a,\DL2}$ is a density variable corresponding to the $-v_{\DL} \leq v_x \leq \infty$ region of velocity space. If the distribution were Maxwellian for all velocities, $n_{a,\DL2}$ would equal the total density $n_{\DL2} \equiv \int_{-\infty}^\infty dv_x\, f_{e,\DL2}$. However, since $f_{e,\DL2} = 0$ for $v_x \leq -v_{\DL}$, $n_{\DL2} < n_{a,\DL2}$. Likewise, $T_e$ is not equal to the total temperature defined from a velocity-space moment of $f_{e}$, but the two are approximately the same as long as $e|\Delta \phi_{\DL}| \gtrsim T_e$. 
 
Putting Eq.~(\ref{eq:fa}) into the current-free condition of Eq.~(\ref{eq:curbal}) provides an expression relating the double layer potential drop and the electron temperature
\begin{equation}
\frac{1}{4} n_{a,\DL2}\, \bar{v}_{e}\, e^{- e |\Delta \phi_{\DL} |/T_{e}} = n_{\DL2}\, e^{-1/2} \sqrt{\frac{T_{e}}{M_i}}  . \label{eq:cabal}
\end{equation}
Here $\bar{v}_{e} \equiv \sqrt{8 T_{e}/(\pi m_e)}$ is an average electron speed. Solving Eq.~(\ref{eq:cabal}) for $|\Delta \phi_{\DL}|$ yields
\begin{equation}
| \Delta \phi_{\DL} |  = - \frac{T_{e}}{e} \ln \biggl( \frac{n_{\DL2}}{n_{a,\DL2}} e^{-1/2} \sqrt{\frac{2 \pi m_e}{M_i}} \biggr)  .
\end{equation}
Recall that $n_{a,\DL2} > n_{\DL2}$, but from the definition $n_{\DL2} \equiv \int_{-\infty}^\infty dv_x\, f_{e,\DL2}$:
\begin{equation}
\frac{n_{\DL2}}{n_{a,\DL2}} = 1 - \frac{1}{2} \textrm{erfc} \sqrt{\frac{e |\Delta \phi_{\DL} |}{T_{e}}}  \approx 1 ,
\end{equation}
since $\erfc (\sqrt{e|\Delta \phi_{\DL}|/T_e}) \sim \mathcal{O}(\sqrt{m_e/M_i}) \ll 1$. 
Thus, the double layer potential drop is approximately the floating potential of a planar probe
\begin{equation}
|\Delta \phi_{\DL}| \approx \frac{T_{e}}{2e} \biggl[ 1 + \ln \biggl( \frac{M_i}{2 \pi m_e} \biggr) \biggr]  . \label{eq:float}
\end{equation}

Equation~(\ref{eq:float}) has previously been derived by Chen\cite{chen:06} in the context of current-free double layers. Although the semi-infinite domain approximation may be useful for a thruster operating in space, it is unable to capture some features of finite laboratory experiments. Equation~(\ref{eq:float}) provides a maximum potential drop that might be expected in the laboratory. Accounting for plasma in a finite downstream expansion chamber leads to some electrons migrating up the double layer and being accelerated into the source region. These electrons fill in part of the otherwise truncated tail of the EVDF. To preserve the current balance in this situation, the double layer potential must be reduced in comparison to Eq.~(\ref{eq:float}) so extra electrons are allowed to leak downstream to balance those coming upstream. This effect will be discussed in more detail in Sec.~\ref{sec:pmax}. 

\subsection{Upstream wall effects: $p_{\min}$\label{sec:pmin}}

\begin{figure}
\includegraphics{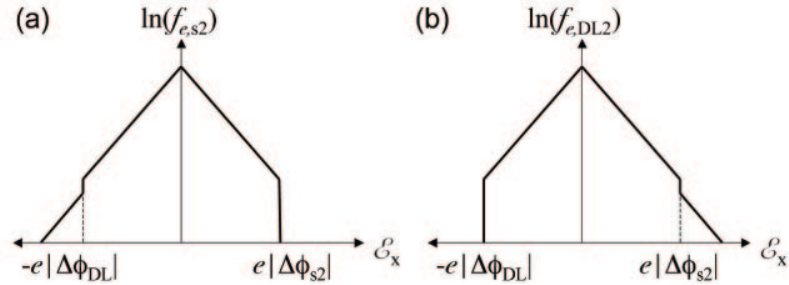}
\caption{Sketch of the natural log of the EVDF for the finite source region of Sec.~\ref{sec:pmin} at (a) the sheath edge of the source wall (P=$s2$), and (b) just upstream of the double layer (P=DL2).}
\label{fg:feb_sketch}
\end{figure}

Next, we consider a geometry with the same semi-infinite and collisionless downstream region as Sec.~\ref{sec:chen}, but allow for a source chamber of finite length. For this case, we model the EVDF at position P as
\begin{eqnarray}
f_{x,\p} = \frac{e^{-v_x^2/v_{Te}^2}}{\sqrt{\pi} v_{Te}}  \label{eq:fmin}
\left\lbrace \begin{array}{cc}
n_{b,\p}, & v_x < -v_{\DL} \\
n_{a,\p},  & -v_{\DL} \leq v_x \leq v_{s2} \\
n_{c,\p},  & v_{s2} < v_x
\end{array} \right. ,
\end{eqnarray}
in which $v_{s2} \equiv \sqrt{2 e |\Delta \phi_{s2}|/m_e}$. The distribution of Eq.~(\ref{eq:fmin}) is shown schematically for positions P=s2 and P=DL2 in Fig.~\ref{fg:feb_sketch}. The distribution is Maxwellian with density $n_{a}$ in the velocity-space interval where particles are confined: $-v_{\DL} \leq v_x \leq v_{s2}$. Outside of this interval (in the tails) the EVDF is depleted from the nominal Maxwellian distribution due to losses to the wall through the upstream sheath, or to the downstream vacuum through the double layer. These regions get repleted in the source primarily due to elastic collisions from the perpendicular to parallel direction. Equation~(\ref{eq:fmin}) models these tail regions by assigning a different density ($n_{b}$ or $n_c$) to the tail regions. It is assumed that these regions can be described by the same temperature as the bulk interval. We also assume that the upstream sheath and double layer are sufficiently thin that they are approximately collisionless. Thus, $n_{c,s2} = n_{b,\DL2} \approx 0$. 

With the assumed boundary conditions, the source chamber is essentially a plane symmetric discharge. Due to this symmetry $n_{b,s2} = n_{c,\DL2}$, which implies $\Delta \phi_{\DL} = \Delta \phi_{s2}$. Applying these assumptions, and putting Eq.~(\ref{eq:fmin}) into Eq.~(\ref{eq:curbal}), yields
\begin{equation}
|\Delta \phi_{\DL}| = - \frac{T_e}{e} \ln \biggl( \frac{n_{\DL2}}{n_{c,\DL2}} e^{-1/2} \sqrt{\frac{2\pi m_e}{M_i}} \biggr) .  \label{eq:pdlb}
\end{equation}
Aside from the density ratio, $n_{\DL2}/n_{c,\DL2}$, Eq.~(\ref{eq:pdlb}) is simply the floating potential from Eq.~(\ref{eq:float}). However, since $n_{\DL2}/n_{c,\DL2} \approx n_{a,\DL2}/n_{c,\DL2} >1$, the extra term acts to reduce the double layer potential. Equation~(\ref{eq:pdlb}) has a viable solution only if 
\begin{equation}
0 < \frac{n_{\DL2}}{n_{c,\DL2}} e^{-1/2} \sqrt{\frac{2\pi m_e}{M_i}} < 1  . \label{eq:pmincon}
\end{equation}
Equation~(\ref{eq:pmincon}) shows that if the there is not enough scattering in the source region, the discharge cannot be maintained. Scattering in the source causes the otherwise missing tails of the EVDF to be filled in, so $n_{c,\DL2}/n_{\DL2} = f(\lambda_{e-n}/L_\textrm{s})$ in which $\lambda_{e-n}$ is the electron-neutral scattering length and $L_\textrm{s}$ is the length of the source region. The particular functional dependence of this relationship depends on details of the scattering cross sections. However, if we assume that it has a simple linear dependence
\begin{eqnarray}
\frac{n_{c,\DL2}}{n_{\DL2}} \approx  \label{eq:linear}
\left\lbrace \begin{array}{cc}
L_{\textrm{s}}/\lambda_{e-n},  & L_{\textrm{s}} < \lambda_{e,n} \\
1,  & L_{\textrm{s}} \geq \lambda_{e,n}
\end{array} \right. 
\end{eqnarray}
this can be used to estimate the minimum neutral pressure required to maintain the discharge. Using $\lambda_{e-n} = 1/(n_n \sigma_{e-n})$ and $n_n = n_o p$, in which $n_o = 3.3 \times 10^{19}$ [m$^{-3}$ mTorr$^{-1}$] and $p$ is in mTorr, Eq.~(\ref{eq:pmincon}) implies
\begin{equation}
p_{\min} \approx \frac{e^{-1/2} \sqrt{2\pi m_e/ M_i}}{n_o \sigma_{e-n} L_{\textrm{s}}}  . \label{eq:pmin}
\end{equation}
For neutral pressures less than Eq.~(\ref{eq:pmin}), a current free double layer is not predicted to be a steady-state solution. 


\subsection{Downstream wall effects: $p_{\max}$\label{sec:pmax}}

\begin{figure}
\includegraphics{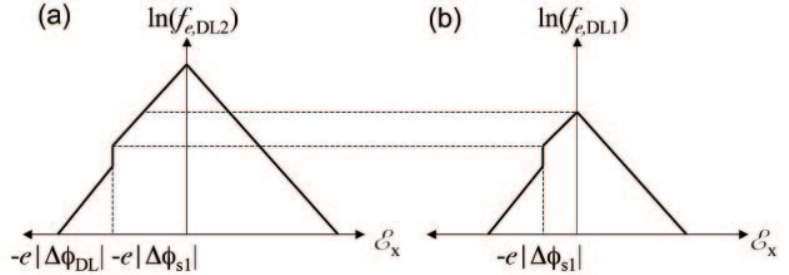}
\caption{Sketch of the natural log of the EVDF for the finite expansion chamber of Sec.~\ref{sec:pmax} at positions (a) just upstream of the double layer (P=DL2), and (b) just downstream of the double layer (P=DL1).}
\label{fg:fec_sketch}
\end{figure}

If the expansion chamber downstream is finite in extent, the sheath at the downstream wall will reflect a population of electrons that can migrate back to the double layer. These are subsequently accelerated into the source chamber. In addition, scattering in the downstream region can partially replete the velocity space interval beyond the downstream sheath cut-off: $v_{s1} \equiv \sqrt{2 e |\Delta \phi_{s1}|/m_e}$. The EVDF just up and downstream of the double layer is shown in Fig.~\ref{fg:fec_sketch} for this case. At the upstream position, the EVDF takes the form
\begin{eqnarray}
f_{x,\DL2} = \frac{e^{-v_x^2/v_{Te}^2}}{\sqrt{\pi} v_{Te}}  \label{eq:fmax}
\left\lbrace \begin{array}{cc}
n_{d,\DL2}, & v_x < -v_{\DL+s1} \\
n_{a,\DL2},  & -v_{\DL+s1} \leq v_x 
\end{array} \right. ,
\end{eqnarray}
in which $v_{\DL+s1} \equiv \sqrt{2e(|\Delta \phi_{\DL}| + |\Delta \phi_{s1}|)/m_e}$. The EVDF just downstream has the same form, but with $v_{\DL+s1}$ replaced by $v_{s1}$. 

Putting the EVDF from Eq.~(\ref{eq:fmax}) into the current-free condition of Eq.~(\ref{eq:curbal}) yields
\begin{equation}
|\phi_{2}| = - \frac{T_e}{e} \ln \biggl( \frac{n_{\DL2} e^{-1/2} \sqrt{2\pi m_e/M_i}}{n_{a, \DL2} - n_{d,\DL2}} \biggr) .  \label{eq:phimax}
\end{equation}
in which $|\phi_2| \equiv |\Delta \phi_{\DL} | + |\Delta \phi_{s1}|$. Equation~(\ref{eq:phimax}) has a viable solution only if
\begin{equation}
0 < \frac{n_{\DL2}}{n_{a,\DL2} - n_{d,\DL2}} e^{-1/2} \sqrt{\frac{2\pi m_e}{M_i}} < 1  .  \label{eq:pmaxcond}
\end{equation}
Assuming $n_{a,\DL2} \approx n_{\DL2}$, Eq.~(\ref{eq:pmaxcond}) requires 
\begin{equation}
\frac{n_{d,\DL2}}{n_{\DL2}} \leq 1 - e^{-1/2} \sqrt{\frac{2\pi m_e}{M_i}} \approx 1.  \label{eq:pmaxy}
\end{equation}
As in Sec.~\ref{sec:pmin}, the precise functional dependence of $n_{d,\DL2}/n_{\DL2}$ due to scattering in the downstream region is difficult to determine. We again assume a simple linear form
\begin{eqnarray}
\frac{n_{d,\DL2}}{n_{\DL2}} \approx  \label{eq:nd}
\left\lbrace \begin{array}{cc}
L_{\textrm{d}}/\lambda_{e-n},  & L_{\textrm{d}} < \lambda_{e,n} \\
1,  & L_{\textrm{d}} \geq \lambda_{e,n}
\end{array} \right. ,
\end{eqnarray}
in which $L_{\textrm{d}}$ is the length of the downstream region. Applying the relations $\lambda_{e-n} = 1/(n_n \sigma_{e-n})$ and $n_n = n_o p$, in which $n_o = 3.3 \times 10^{19}$ [m$^{-3}$ mTorr$^{-1}$] and $p$ is in mTorr, Eqs.~(\ref{eq:pmaxy}) and (\ref{eq:nd}) provide an estimate for the maximum neutral pressure the current-free double layer solution can support
\begin{equation}
p_{\max} \approx \frac{1}{n_o \sigma_{e-n} L_{d}}  . \label{eq:pmax}
\end{equation}
Equation~(\ref{eq:pmax}) shows that when too many electrons migrate up the double layer from downstream, the double layer potential cannot adjust enough to preserve current balance. The physics justification for Eqs.~(\ref{eq:pmin}) and (\ref{eq:pmax}) are the same as those determining the $p_{\min}$ and $p_{\max}$ in Ref.~\onlinecite{lieb:06}. However, the analysis is different since Ref.~\onlinecite{lieb:06} is based on a 3D fluid model which is diffusion dominated, while this is a 1D kinetic model where collisions are modeled with the simple linear estimates of Eqs.~(\ref{eq:linear}) or (\ref{eq:nd}). 

\subsection{Finite 1D domain\label{sec:full}}

\begin{figure}
\includegraphics{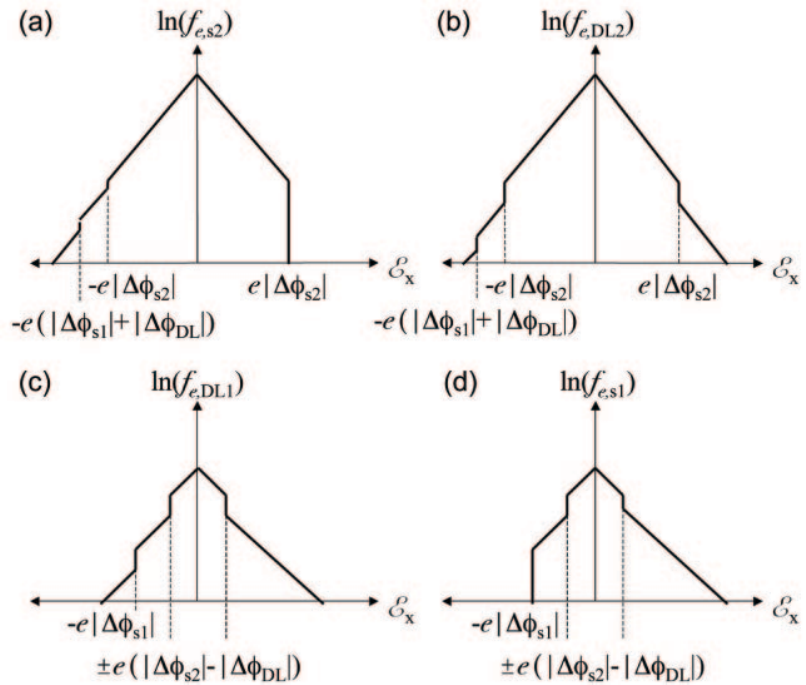}
\caption{Sketch of the natural log of the EVDF for the finite 1D domain of Sec.~\ref{sec:full} at (a) the sheath edge of the source wall (P=$s2$), (b) just upstream of the double layer (P=DL2), (c) just downstream of the double layer (P=DL1), and (c) the sheath edge of the expansion chamber wall (P=$s1$).}
\label{fg:fed_sketch}
\end{figure}

The full simulation domain has two boundaries and the length of both the source and downstream domains can be comparable to $\lambda_{e-n}$ (depending on the neutral pressure). For low neutral pressures, we expect that the EVDF will reflect features of losses to both walls in the manner depicted in Fig.~\ref{fg:fed_sketch}. Figure~\ref{fg:fed_sketch} shows a sketch of the expected EVDF at four locations in the simulation domain: the upstream sheath edge (s2), just upstream of the double layer (DL2), just downstream of the double layer (DL1), and the downstream sheath edge (s1). As the figure demonstrates, this model of depletion due to wall losses and repletion due to scattering predicts several features of the EVDF that can be tested in the stimulations. At low neutral pressures, particularly, these features should be clearly visible and their location in velocity-space can be compared with the predicted values dependent on the sheath and double layer potential drops. As the neutral pressure is increased, repletion becomes more prevalent and velocity-space intervals affected by wall losses are more quickly filled in. At higher neutral pressures, it is expected that the depleted intervals become more difficult to distinguish until finally the downstream region becomes too collisional to support the current-free double layer solution. 

The sheath and double layer potential drops can be written in terms of the densities of the various intervals in velocity space, in a similar manner to Secs.~\ref{sec:pmin} and \ref{sec:pmax}, but the extra velocity-space intervals significantly complicate the analysis. The only qualitative difference to the analysis of the previous two sections is that accounting for migration of a small current of downstream electrons into the upstream region leads to a slight asymmetry in the source region (so $|\Delta \phi_{s2}| \approx |\Delta \phi_{\DL2}|$, instead of $|\Delta \phi_{s2}| = |\Delta \phi_{\DL2}|$). We expect that Eqs.~(\ref{eq:pmin}) and (\ref{eq:pmax}) remain good approximations for the minimum and maximum pressure limits, and that the double layer potential drop remains close to the floating potential of Eq.~(\ref{eq:float}) for intermediate pressures. These estimates will be compared with simulation data in Sec.~\ref{sec:sims}. 

\section{Description of the Phoenix code\label{sec:code}}

{\sc phoenix} is a PIC-MCC code that is 1D in space and 3D in velocity phase-space (1D-3V). It is designed to be identical to the \emph{JanuS} code described in Meige {\it et al}.\cite{meig:05} The left wall (source chamber) is a floating boundary, which is achieved computationally by inserting a capacitor there. The right wall (expansion chamber) is conducting, which is implemented by removing all particles that reach the cell defining that boundary. Collisions between macroparticles (typically representing $\sim 10^9$ real particles) are simulated with a Monte Carlo technique including the null collision method based on the algorithm developed by Vahedi and Surendra.\cite{vahe:95} The gas species here is argon. The cross section for electron impact ionization was taken from Krishnakumar and Srivastava,\cite{kris:88} and electron excitation collisions from de Heer \textit{et al}.\cite{dehe:79} The electron-argon elastic scattering cross sections were taken from Ferch \textit{et al}\cite{ferc:85} for 0-20 eV and from de Heer \textit{et al}\cite{dehe:79} for 20-3000 eV. These are also collected in Hayashi.\cite{haya:82} The cross sections for argon ion charge-exchange and ionization collisions are from Phelps.\cite{phel:94} 

%

The plasma is generated by first loading a small number of macroparticles (typically 1000) with a spatially uniform Maxwellian distribution of temperature 1 eV throughout the simulation domain. Electrons are heated in a single Cartesian velocity-space direction ($\hat{y}$) perpendicular to the spatial dimension ($\hat{x}$) using the inductive heating method described in Meige {\it et al}.\cite{meig:05} The macroparticle density initially increases due to electron-neutral ionization collisions.  Eventually, a steady-state is reached where particle generation balances particle loss. This typically occurs within $25\, \mu$s and the typical time step used is $50$ ps.  The number of macroparticles in steady-state is $\gtrsim 10^5$. The parameters used in all simulations are summarized in Table~\ref{tb:params}, except that the neutral pressure was varied for the simulations shown in Figs.~\ref{fg:EVDFx_quad}, \ref{fg:phi_pres} and \ref{fg:phi_p}. The $q_{\textrm{factor}}$ was also adjusted for these to meet the $\gtrsim 10^5$ macroparticle condition. These calculations were performed on a desktop PC, and each run took 2-5 days. 


\begin{table}
\begin{tabular}{l*{6}{l}r}
\hline \hline
Quantity         		    & Value  \\
\hline
Neutral pressure	  & 1 mTorr   \\
Domain length         	   & 10 cm    \\
Number of grid cells           & 250   \\
Time step 		     & $5 \times 10^{-11}$ s   \\
Total run time			& 25 $\mu$s \\
Antenna frequency ($\omega_o/2\pi$)       & 10 MHz \\
Antenna current density amplitude	&100 A/m$^2$ \\
$q_{\textrm{factor}}$ 			& $8 \times 10^8$ \\
$\nu_{\textrm{max}}$ 		& $1 \times 10^6$ s$^{-1}$ \\
\hline \hline
\end{tabular}
\caption{Parameters used for all simulations except those shown in Figs.~\ref{fg:EVDFx_quad}, \ref{fg:phi_pres} and \ref{fg:phi_p}, in which the neutral pressure was changed. The $q_{\textrm{factor}}$ was also adjusted for these so that the total number of particles in steady-state exceeded $10^5$.}
\label{tb:params}
\end{table}

In the experiments, a double layer forms due to the expansion of the plasma volume. As the volume expands, the plasma density drops. If this density drop is steep enough, a double layer will form. Since the simulations have only one spatial dimension, volume expansion cannot be simulated self-consistently. Instead, a density drop is imposed by removing particles from the system at a set frequency defined by a profile and amplitude. In Ref.~\onlinecite{meig:05}, various linear loss profiles were used to generate a double layer, but these did not necessarily represent the effective loss profile associated with an expanding magnetic field. Here we modify the loss profile to more closely resemble a diverging solenoidal magnetic field. 

\begin{figure}
\includegraphics{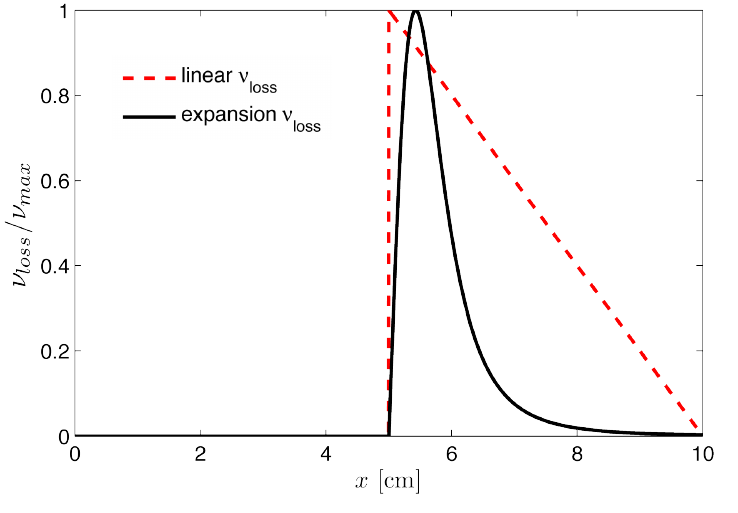}
\caption{Loss profiles implemented in {\sc phoenix} to simulate plasma volume expansion downstream. The triangular loss profile (dashed red line) was used in Ref.~\onlinecite{meig:05} and the curve representing an expanding magnetic field (solid black line) is from Eq.~(\ref{eq:mexpand}).}
\label{fg:nu_profile}
\end{figure}

The vacuum magnetic field on axis from the coil closest to the expansion chamber is $B_o [1+(x-x_c)^2/R^2]^{-3/2}$, in which $R$ is the coil radius, $x_\textrm{c}$ is the axial position of the coil and $B_o \equiv \mu_o I/(2R)$ where $I$ is the coil current. We assume that the magnetic field is constant inside the source chamber, so the magnetic field on axis throughout the domain is 
\begin{eqnarray}
B(x) = B_o  \label{eq:bfield}
\left\lbrace \begin{array}{cc}
1, & 0 \leq x \leq x_{\textrm{c}} \\
(1 + X^2)^{-3/2}, & x_{\textrm{c}} \leq x \leq L
\end{array} \right. 
\end{eqnarray}
in which $X\equiv (x-x_c)/R$. The volume expansion obeys $V/V_o = (r/r_o)^2 = B/B_o$,\cite{chen:06} so the change in volume satisfies $V_o^{-1} dV/dx = B_o^{-1} |dB/dx|$.  Thus, an appropriate loss profile for magnetic field expansion has the form $\nu_{\textrm{loss}} \approx (v/B_o) |d B/dx|$, in which $v$ is some characteristic velocity. For the field of Eq.~(\ref{eq:bfield}), the loss profile is
\begin{eqnarray}
\nu_\textrm{loss}(x) = 
\left\lbrace \begin{array}{cc}  \label{eq:mexpand}
0, & 0 \leq x \leq x_{\textrm{c}} \\
3 \nu_o X(1+X^2)^{-5/2}, & x_{\textrm{c}} \leq x \leq L
\end{array} \right. 
\end{eqnarray}
in which $\nu_o \equiv v/R$. Note that $\nu_{\max} \approx 0.86 \nu_o$. For all our simulations we chose $L = 10$ cm and $x_c = L_s = 5$ cm. In the experiments, $R/L_s \simeq 0.17$, and in order to preserve this ratio we take $R=1.7$ cm in the simulations. We will also choose $\nu_o = 1 \times 10^6$ s$^{-1}$, which corresponds to 1 eV electrons (the initial electron temperature). The loss profile of Eq.~(\ref{eq:mexpand}) is shown in Fig.~\ref{fg:nu_profile}, along with the linear loss profile used in Ref.~\onlinecite{meig:05}. Unless otherwise specified, the simulation results presented in the following sections used the loss profile from Eq.~(\ref{eq:mexpand}). 

Although this simplified simulation geometry can provide insight into the mechanisms of double layer formation and the role of the EVDF, especially in testing the model of Sec.~\ref{sec:basic}, it is not a quantitatively accurate model of the experiments. Since the code is 1D, it does not capture radial effects that have been the topic of recent experimental work.\cite{corr:08,taka:08,taka:10c} Also, the 10 cm length of the simulation domain is nearly an order of magnitude shorter than the axial length of the experiments.\cite{char:03,cohe:03}  Aside from these geometrical effects, one also needs to be cognizant of the physics limitations of this model when interpreting the simulation data. The loss profile is a mock-up of the density drop due to an expanding field, but there is no actual magnetic field in the simulations. For instance, $\nabla B$ drifts may play a role in the expansion region, but are not captured in the simulations. Since the loss profile removes particles randomly (independent of energy), slower particles are more likely to be removed in the loss region. Also, the neutral density is assumed to be uniform and constant, so effects of neutral depletion, which may be important in experiments,\cite{fruc:06} are not captured. 



\section{Simulation results\label{sec:sims}}

\begin{figure}
\includegraphics{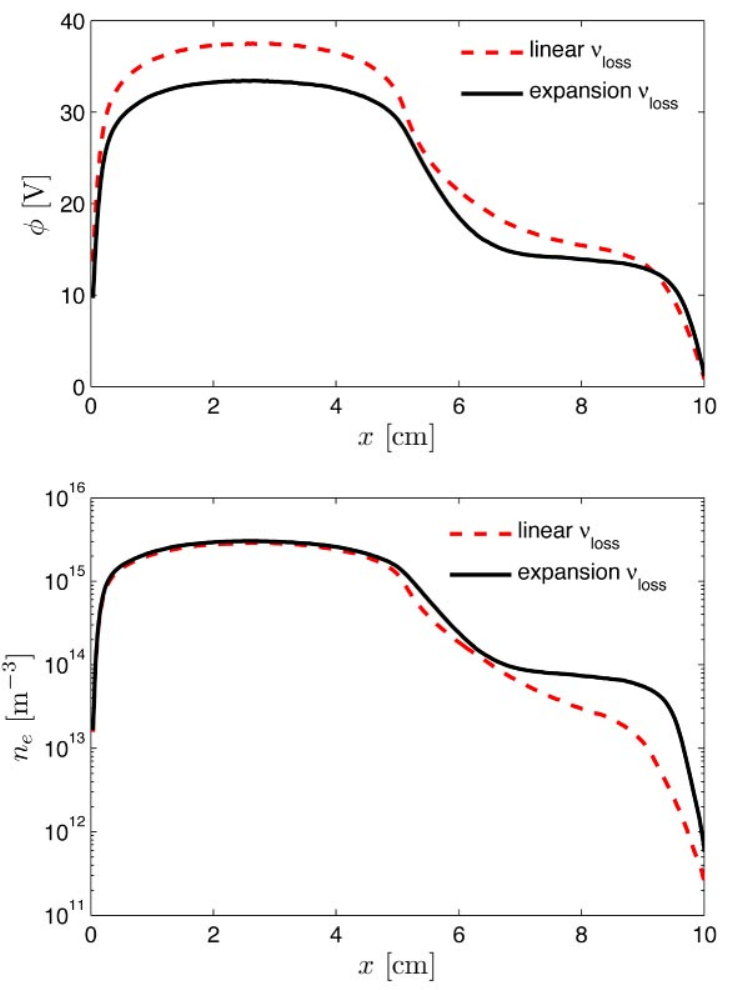}
\caption{Plasma potential and electron density throughout the simulation domain. The red dashed line corresponds to a simulation that used the linear expansion profile and the solid black line to one that used the expanding magnetic field profile; see Fig.~\ref{fg:nu_profile}.}
\label{fg:phi_prof}
\end{figure}

\begin{figure}
\includegraphics{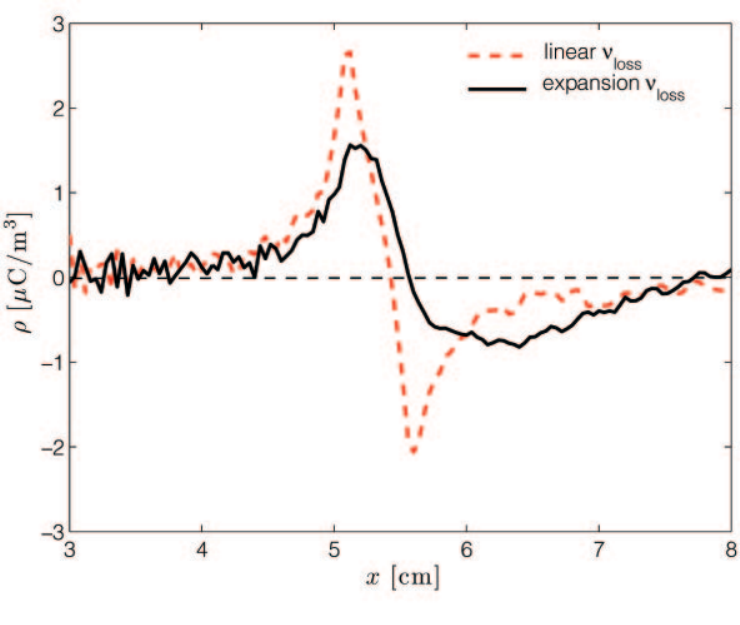}
\caption{Charge density as a function of position showing the adjacent regions of positive and negative space charge that define a double layer. The red dashed line corresponds to a simulation that used the linear loss profile and the black solid line to one that used Eq.~(\ref{eq:mexpand}).}
\label{fg:rho}
\end{figure}

The electrostatic potential and density are shown in Fig.~\ref{fg:phi_prof} for both the linear and expanding magnetic field loss profiles from Fig.~\ref{fg:nu_profile}. The data shown throughout this work was averaged over a few rf periods.  The density and potential profiles are qualitatively similar for either loss profile. However, the upstream potential is a few volts less for the magnetic field expansion profile [from Eq.~(\ref{eq:mexpand})].  Also, the double layer potential drop is steeper and the downstream region more uniform for Eq.~(\ref{eq:mexpand}). These are due to the relative narrowness of the magnetic field expansion profile, which is shown in Fig.~\ref{fg:nu_profile}. The characteristic step potential profile of a double layer is seen in Fig.~\ref{fg:phi_prof}. Figure~\ref{fg:rho} shows a profile of the charge density: $\rho = e(n_i-n_e)$. It has been suggested in previous literature that the potential profile of expanding plasmas, which are usually deemed ``double layers,'' are actually single layers similar to sheaths.\cite{chen:06} Figure~\ref{fg:rho} shows explicitly adjacent regions of positive and negative space charge, which is typically the property used to define a double layer.\cite{hers:85} Thus, we conclude that double layers, not single layers, are found in these simulations. 

\subsection{Ion beams and the IVDF}

\begin{figure}
\includegraphics{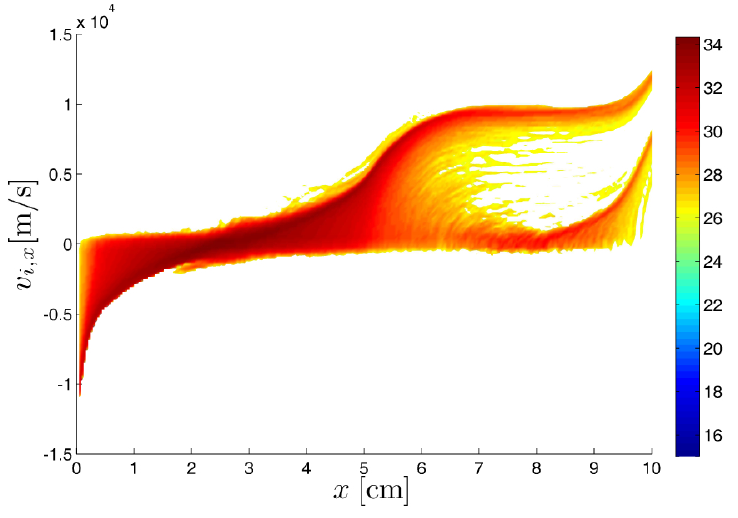}
\caption{A color map showing the natural logarithm of the ion velocity distribution in the $\hat{x}$ direction throughout the simulation domain. Colors corresponding to higher numbers on the color bar represent higher concentration of particles. }
\label{fg:IVDFx}
\end{figure}

\begin{figure}
\includegraphics{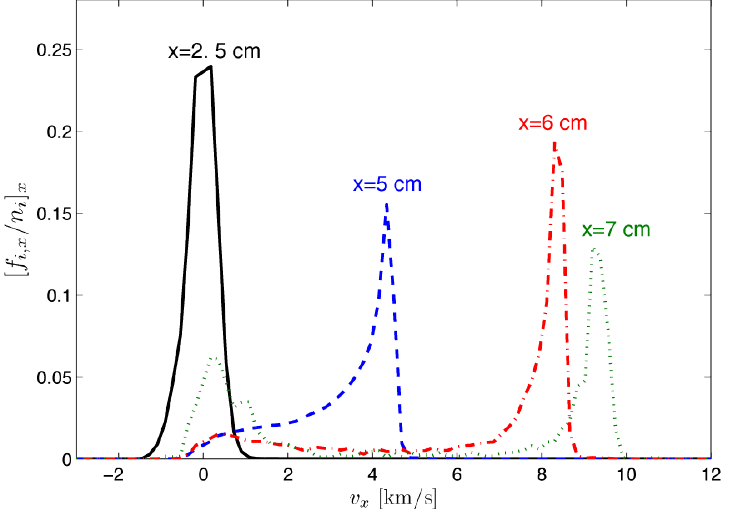}
\caption{The ion velocity distribution function in the $\hat{x}$ direction divided by the ion density at the axial locations $x=$2.5 cm, 5 cm, 6 cm, and 7 cm. These correspond to the center of the source chamber, middle of the double layer, just downstream of the double layer and middle of the downstream region.  }
\label{fg:IVDFx_cuts}
\end{figure}

The ion velocity distribution function (IVDF) in the $\hat{x}$ direction is shown as a color map in Fig.~\ref{fg:IVDFx} throughout the simulation domain. In the central source region, it is a stationary Maxwellian. Ions are accelerated by the sheath electric fields at each boundary, so the IVDF has a flow shift there and a lower energy tail due to ion scattering.  A supersonic ion beam is generated by the double layer potential drop and this beam is maintained at a constant speed downstream (until the downstream sheath is reached). The beam speed is approximately $1 \times 10^4$ m/s. In the next section it will be shown that $T_e \approx 4$ eV downstream, so this beam travels at $\approx 3 c_s$. This agrees with the expected flow speed if the double layer potential drop is the floating potential of Eq.~(\ref{eq:float}): $V_i \simeq \sqrt{2 e |\Delta \phi_{\DL}|/M_i} = 3.1 c_s$ (for argon). One-dimensional cuts of the beam distribution are shown in Fig.~\ref{fg:IVDFx_cuts}. The largest ion-neutral cross section at the ion beam energies is charge exchange. This expectation is corroborated by the data of Figs.~\ref{fg:IVDFx} and \ref{fg:IVDFx_cuts}, which show that ions lost from the beam show up directly as low-energy thermal particles. If the collisions were elastic, the beam would slow gradually, which does not happen. The ion beams shown in Figs.~\ref{fg:IVDFx} and \ref{fg:IVDFx_cuts} agree with the previous simulations,\cite{meig:05} and the $\sim 3 c_s$ speed downstream agrees with previous measurements.\cite{sun:05,kees:05}

\subsection{EVDFs and electron temperature}

\begin{figure}
\includegraphics{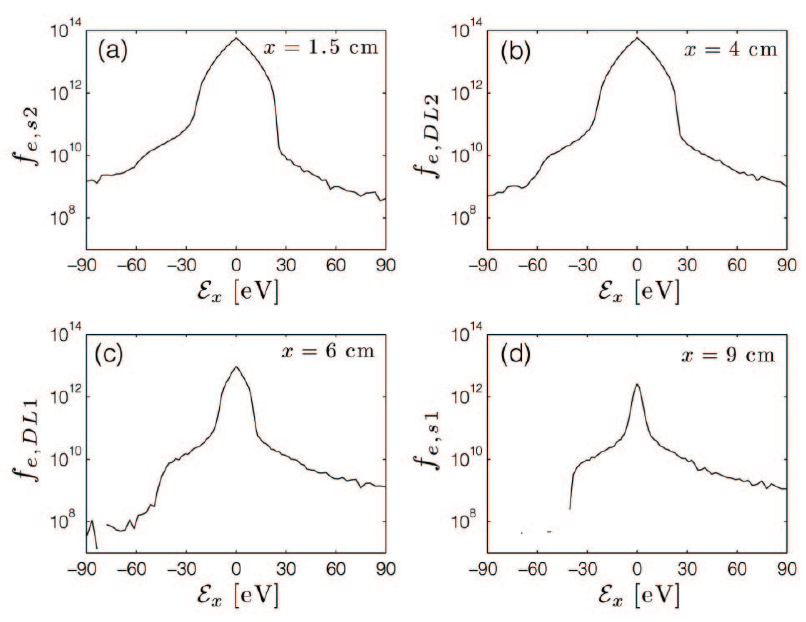}
\caption{The $\hat{x}$-directed EVDF for a 0.1 mTorr neutral pressure simulation at (a) the source sheath edge ($x=1.5$ cm), (b) just upstream of the double layer ($x=4$ cm), (c) just downstream of the double layer ($x=6$ cm), and (d) the expansion chamber sheath edge ($x=9$ cm).}
\label{fg:EVDFx_quad}
\end{figure}

\begin{figure}
\includegraphics{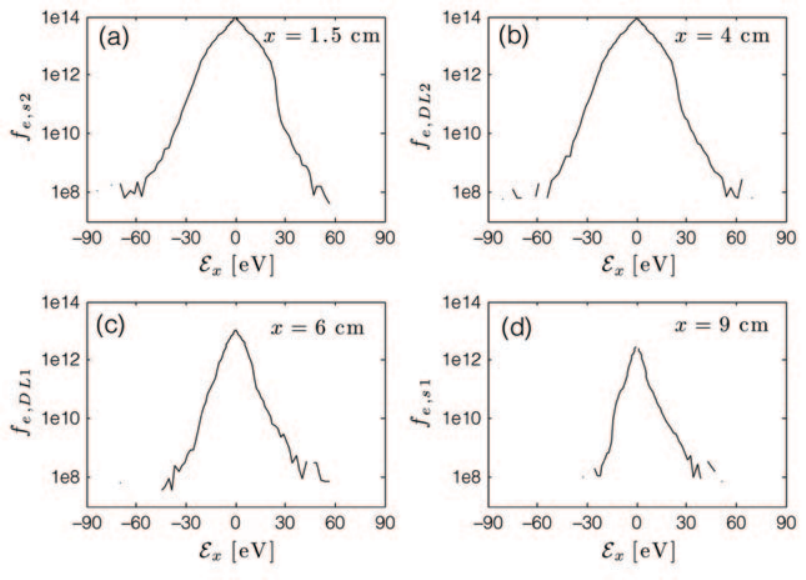}
\caption{The $\hat{x}$-directed EVDF for a 1 mTorr neutral pressure simulation at (a) the source sheath edge ($x=1.5$ cm), (b) just upstream of the double layer ($x=4$ cm), (c) just downstream of the double layer ($x=6$ cm), and (d) the expansion chamber sheath edge ($x=9$ cm).}
\label{fg:EVDFx_quad2}
\end{figure}

The EVDF in the $\hat{x}$ direction is shown in Figs.~\ref{fg:EVDFx_quad} and \ref{fg:EVDFx_quad2} for neutral pressures of 0.1 and 1 mTorr. In each figure, the EVDF is shown at four positions in the simulation domain: the source region sheath edge (s2=1.5 cm), just upstream of the  double layer (DL2=4 cm), just downstream of the double layer (DL1=6 cm), and the expansion region sheath edge (s1=9 cm). These figures can be compared with the model predictions from Fig.~\ref{fg:fed_sketch} of Sec.~\ref{sec:full}. 

For low neutral pressure (0.1 mTorr), each of the features predicted in Fig.~\ref{fg:fed_sketch} of Sec.~\ref{sec:full} can be seen in the simulation data of Fig.~\ref{fg:EVDFx_quad}. Here, the potential drop of the source sheath is $|\Delta \phi_{s2}| = 23$ V, the double layer is $|\Delta \phi_{\DL}| = 18$ V, and the expansion region sheath is $|\Delta \phi_{s1}| = 40$ V.  At the source region sheath edge, the distribution is depleted from the nominal Maxwellian for $\mathcal{E}_x > e|\Delta \phi_s|$ by more than two orders of magnitude. This is the truncation due to electron loss to the source boundary that was predicted in Sec.~\ref{sec:full}. The EVDF is also depleted for $\mathcal{E}_x < - e|\Delta \phi_{s2}|$ due to the same wall losses, but it has been partially repleted due to scattering over the whole the simulation domain. Figure~\ref{fg:EVDFx_quad} also shows additional depletion for $\mathcal{E}_x < -e(|\Delta \phi_{s1}| + |\Delta \phi_{\DL}|) = 58\ \textrm{eV}$ due to losses to the expansion chamber boundary. Likewise, the predicted features of the EVDF at each of the other positions ($x=\DL2, \DL1,$ and $s1$) compare well with the predictions from Fig.~\ref{fg:fed_sketch}. 

As the neutral pressure is increased, Fig.~\ref{fg:EVDFx_quad2} shows that repletion of the velocity space intervals subject to wall losses also increases. This is simply due to the increase in electron-neutral scattering that occurs for higher neutral density. The dominant scattering processes for electrons on the tail of the Maxwellian (beyond the sheath energy) are elastic and ionization collisions. The elastic processes cause incident electrons to change velocity by a small amount during each scattering event. Repletion of the tail happens from a combination of high energy electrons scattering from the perpendicular to parallel direction and electrons in the parallel direction gaining energy from several scattering events. 

\begin{figure}
\includegraphics{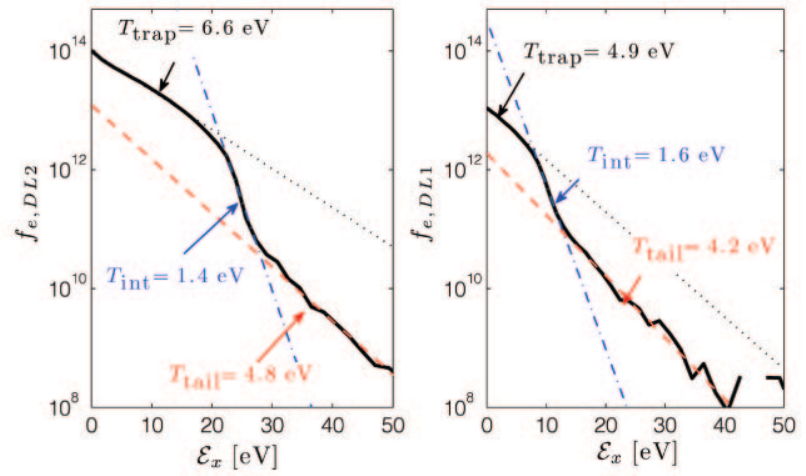}
\caption{EVDF in the $+\hat{x}$ direction at the upstream edge ($x=\DL2$) and downstream edge ($x=\DL1$) of the double layer. Shown are linear least squares fits to three intervals of velocity space: trapped, tail and intermediate. Also shown are the effective temperatures of each interval. This simulation was run with a 1 mTorr neutral pressure.}
\label{fg:fit1}
\end{figure}

Figure~\ref{fg:fit1} shows the $+\hat{x}$ direction of the EVDF from Fig.~\ref{fg:EVDFx_quad2} at positions $\DL2$ and $\DL1$. Three populations of electrons are present. These include the trapped electrons below the break energy and tail electrons past the break energy that were included in the models of Sec.~\ref{sec:basic}. The step from one population to the other, which was assumed to be a sharp step in the model, is broadened due to scattering. Electrons in this intermediate velocity-space interval form a third population. Figure~\ref{fg:fit1} also shows the effective temperature of each of these three intervals. These effective temperatures are calculated using a linear least squares fit to the data in the form $\ln (f/f_o) = A \mathcal{E}_x + B$, in which $f$ is the simulation data for the EVDF in the $\hat{x}$ direction, $f_o = f(v_x=0)$ and $A$ and $B$ are the constants determined from the linear least squares fit. Assuming each interval is close to Maxwellian, i.e., straight lines in Fig.~\ref{fg:fit1}, the effective temperature for that interval is $T = 1/|A|$. Although the step between the trapped and tail populations is not immediate, the temperature characterizing this interval is much colder than either the trapped or tail temperature. The model of Sec.~\ref{sec:basic} effectively assumes $T_{\textrm{int}} = 0$. 

In the models of Sec.~\ref{sec:basic}, it was assumed that both the trapped and tail populations had the same effective temperature $T_e$. Figure~\ref{fg:fit1} suggests that this is a reasonable assumption. However, the intermediate population was not included in the model and presents a complication in that these electrons are effectively colder.  In particular, we want to determine what temperature should be used in calculating the double layer potential drop. Upstream, most electrons are trapped so we expect the total temperature there to be approximately the temperature of the trapped population. However, most of the trapped electrons do not contribute to the current balance (see Sec.~\ref{sec:pmin}). Only electrons that have enough energy to escape the double layer $\mathcal{E}_x > |\Delta \phi_{\DL}|$, i.e., those that make it downstream, contribute. Thus, we expect that the appropriate temperature to use in calculating the double layer potential drop should be the downstream temperature. This includes a small part of the trapped population [$0 \leq \mathcal{E}_x \lesssim e( |\Delta \phi_{s2}| - |\Delta \phi_{\DL}|)$], the whole intermediate population, and the whole tail population [$\mathcal{E}_x \gtrsim e( |\Delta \phi_{s2}| - |\Delta \phi_{\DL}|)$]. As long as the neutral pressure is within the range that a double layer can form, the double layer potential drop is expected to be approximately the floating potential in which the temperature is the downstream temperature:
\begin{equation}
|\Delta \phi_{\DL}| \simeq \frac{T_{e,\textrm{dn}}}{2 e} \biggl[ 1 + \ln \biggl( \frac{M_i}{2\pi m_e} \biggr) \biggr] .  \label{eq:dldrop}
\end{equation}

\begin{figure}
\includegraphics{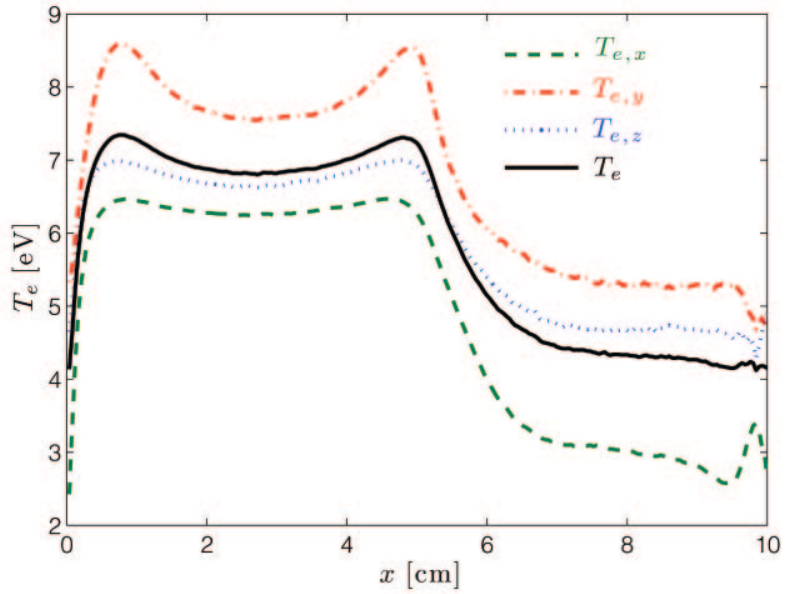}
\caption{Electron temperature ($T_e$) calculated from the EVDF using the moment definition of Eq.~(\ref{eq:temp}) (black solid line). Also shown are effective temperatures in the $\hat{x}$ direction (green dashed line), $\hat{y}$ direction (red dash-dotted line) and $\hat{z}$ direction (blue dotted line) calculated using Eq.~(\ref{eq:tx}).  }
\label{fg:Te}
\end{figure}

Figure~\ref{fg:Te} shows the electron temperature throughout the simulation domain. This is calculated from the EVDF using the moment definition:
\begin{equation}
T \equiv \frac{1}{3} \frac{m}{n} \int d^3 v\, (\vc{v} - \vc{V})^2 f \label{eq:temp}
\end{equation}
in which $\vc{V} \equiv \frac{1}{n} \int d^3v\, \vc{v}\, f  \label{eq:flow}$ is the fluid flow velocity. Also shown are three characteristic temperatures of the EVDF in each Cartesian direction. Along $\hat{x}$, this directional temperature is defined as
\begin{equation}
T_x \equiv \frac{m}{n} \int d^3v\ (v_x - V_x)^2\, f  \label{eq:tx}
\end{equation}
with analogous definitions for the $\hat{y}$ and $\hat{z}$ directions. The total temperature can be expressed in terms of the directional temperatures with the relation: $T = (T_x + T_y + T_z)/3$. Figure~\ref{fg:Te} shows that electrons are significantly colder in the downstream region than the upstream region. This is because the colder intermediate population is a greater fraction of the total electron density downstream than upstream. Upstream, most of the electrons are in the trapped interval ($-e|\Delta \phi_{\DL}| \leq \mathcal{E}_x \leq e |\Delta \phi_{\DL}|$) and these electrons set the upstream temperature; see Fig.~\ref{fg:Te}. Electrons in the $\hat{x}$ direction are colder than either of the perpendicular directions because the predominant sink for electron energy is wall losses, which only happens in the $\hat{x}$ direction. Electrons are hottest in the $\hat{y}$ direction because this is the only direction that electrons are heated. Figure~\ref{fg:Te} also shows that there is some electron heating from the presheaths of the upstream sheath and double layer. Using the 4 eV downstream electron temperature from Fig.~\ref{fg:Te}, Eq.~(\ref{eq:dldrop}) predicts $|\Delta \phi_{\DL}| \approx 21$ eV. This agrees well with the approximately 20 eV potential drop shown in Fig.~\ref{fg:phi_prof}. 


\subsection{Neutral pressure limits\label{sec:press}}

\begin{figure}
\includegraphics{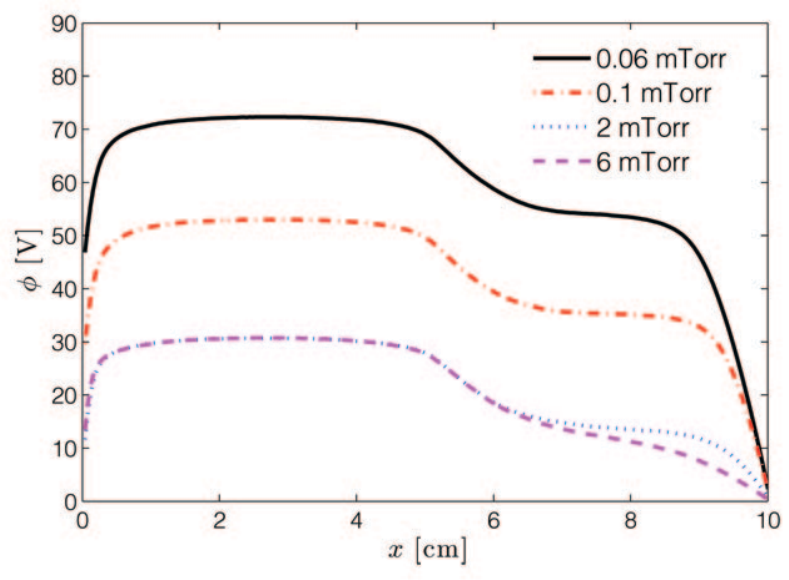}
\caption{Electrostatic potential profile through the simulation domain for various neutral pressures. }
\label{fg:phi_pres}
\end{figure}

\begin{figure}
\includegraphics{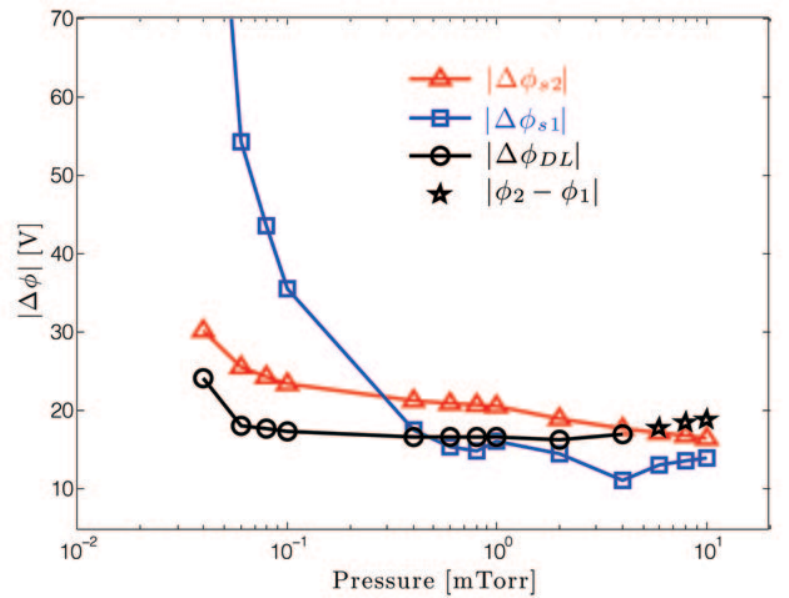}
\caption{Upstream sheath potential drop $|\Delta \phi_{s2}|$ (red triangles), double layer potential drop $|\Delta \phi_{\DL}|$ (black circles) and downstream sheath potential drop $|\Delta \phi_{s1}|$ (blue squares) as a function of neutral pressure. Stars show points for $|\phi_2 - \phi_1|$, but where the downstream electric field is sufficiently strong that it is not considered a double layer solution.}
\label{fg:phi_p}
\end{figure}

The potential profile through the simulation domain is shown in Fig.~\ref{fg:phi_pres} for neutral pressures of 0.06, 0.1, 2 and 6 mTorr. The potential drops $|\Delta \phi_{s2}|$, $|\Delta \phi_{\DL}|$ and $|\Delta \phi_{s1}|$ are also shown in Fig.~\ref{fg:phi_p} for several neutral pressures ranging from 0.04 to 10 mTorr. These were calculated using $|\Delta \phi_{s2}| = \phi_2 - \phi_{sw}$, $|\Delta \phi_{\DL}| = \phi_2 - \phi_1$, and $|\Delta \phi_{s1}| = \phi_1 - \phi_o$ where $\phi_{sw} = \phi (x=0)$, $\phi_2 = \phi (x=2.5\, \textrm{cm})$, $\phi_1 = \phi(x = 7.5\, \textrm{cm})$ and $\phi_{o} = \phi(x = 10\, \textrm{cm}) = 0$.  The figures show that as the neutral pressure is decreased, the downstream sheath drop increases. The upstream sheath and double layer potential remain nearly constant. Simulations were also run at 0.01 and 0.02 mTorr, but no double layer was found. In these cases the plasma density was very low, even though it was stable in time, which is characteristic of there not being enough ionization to sustain the discharge. Thus, the minimum pressure to sustain the discharge in the simulation was in the range between 0.02 and 0.04 mTorr. Although, Fig.~\ref{fg:phi_p} shows that at 0.04 mTorr the downstream sheath potential drop becomes very large (the data point is at 114 V, which is off of the figure) and this does not seem physically reasonable. Thus, maybe the minimum neutral pressure should be considered close to 0.04 mTorr. Figure~\ref{fg:phi_pres} shows that for a neutral pressure of 6 mTorr, the potential profile in the downstream region is no longer flat, but linearly decreases from the double layer to the downstream sheath. This is characteristic of a non-neutral downstream region, and the breakdown of the current-free double layer solution. At 2 mTorr, the potential in the downstream region is flat between the double layer and sheath, suggesting that the double layer solution breaks down between 2 and 6 mTorr.  Data points for $|\phi_2 - \phi_1|$ in this high pressure region, which are not considered double layer solutions, are shown as stars in Fig.~\ref{fg:phi_p}.

Equations~(\ref{eq:pmin}) and (\ref{eq:pmax}) provided predictions for the minimum and maximum neutral pressures that can support a current-free double layer solution. The source and downstream simulation domain lengths are $L_s = L_d = 5$ cm and for thermal ($\approx 4$ eV) electrons, $\sigma_{e-n} \simeq 1 \times 10^{-19}$ m$^{-3}$.\cite{ferc:85,haya:82}  Using these parameters, Eq.~(\ref{eq:pmin}) yields $p_{\min}\approx 0.03$ mTorr and Eq.~(\ref{eq:pmax}) yields $p_{\max} \approx 6$ mTorr. Both of these estimates are consistent with the simulation results of $p_{\min} \simeq 0.04$ mTorr and $p_{\max} \simeq 2-6$ mTorr. 

\section{Summary\label{sec:sum}}

A model for the EVDF in an expanding plasma with a current-free double layer was developed and shown to compare well with results of a PIC simulation. The dominant mechanisms determining the EVDF are depletion of high energy electrons due to boundary losses and repletion of these energy intervals due to scattering. The degree to which these velocity intervals are repleted was shown to depend on the ratio of the electron-neutral collision length to the system size: $\lambda_{e-n}/L$. Assuming a simple linear dependence on this parameter, a model for the range of neutral pressures that can support a double layer was developed. The pressure minimum [Eq.~(\ref{eq:pmin})] is determined by the minimum scattering needed to sustain the discharge. The pressure maximum [Eq.~(\ref{eq:pmax})] is determined by current balance through the double layer. When the neutral pressure is high, abundant electron scattering in the downstream region generates a large flux of electrons that can migrate back to the double layer and be accelerated by it into the upstream region. If too many electrons do this, which happens at high pressure, current balance across the double layer cannot be maintained. The maximum double layer potential drop for this configuration is the floating potential using the downstream electron temperature. Electrons traveling from the downstream to the upstream region causes a slight decrease from the maximum. Although this model and simulation used a 1D domain, the mechanisms of depletion due to wall losses and repletion due to scattering are expected to be similar in the experiments. These results provide information about the EVDF that is essential for the development of a comprehensive analytic model of the experiments. 

\begin{acknowledgments}

One of the authors (SDB) acknowledges the generous hospitality of the SP3 group during his visit to Australia National University over the Australian winter of 2010. This work was supported by the United States National Science Foundation and the Australian Academy of Sciences under East Asia and Pacific Summer Institute (EAPSI) award number 1015362. 

\end{acknowledgments}

\bibliography{refs.bib}

\end{document}